\documentclass[prb,twocolumn,amsmath,amssymb,superscriptaddress]{revtex4-1}
\usepackage{graphicx}
\usepackage{amsthm}
\usepackage{array}
\usepackage{hyperref}
\usepackage{color}
\usepackage{mathtools}
\usepackage{bm,times,enumitem}
\hypersetup{hidelinks}

\newcommand{\secNAT}[1]{\vspace{1ex}\noindent\textbf{\fontsize{11}{11}\selectfont #1}\\}
\newcommand{\secNATend}[1]{\vspace{2ex}\noindent\textbf{\fontsize{11}{11}\selectfont #1}\\}

\begin{document}
\title{High-fidelity three-qubit \textit{i}Toffoli gate for fixed-frequency superconducting qubits}
\author{Yosep Kim}
\thanks{Correspondence to Y.K. (yosep9201@gmail.com)}  
\affiliation{Computational Research Division, Lawrence Berkeley National Laboratory, Berkeley, California 94720, USA}
 
\author{Alexis Morvan}
\affiliation{Computational Research Division, Lawrence Berkeley National Laboratory, Berkeley, California 94720, USA}

\author{Long B. Nguyen}
\affiliation{Computational Research Division, Lawrence Berkeley National Laboratory, Berkeley, California 94720, USA}

\author{Ravi K. Naik}
\affiliation{Computational Research Division, Lawrence Berkeley National Laboratory, Berkeley, California 94720, USA}
\affiliation{Department of Physics, University of California, Berkeley, California 94720, USA}

\author{Christian J\"{u}nger}
\affiliation{Computational Research Division, Lawrence Berkeley National Laboratory, Berkeley, California 94720, USA}

\author{Larry Chen}
\affiliation{Department of Physics, University of California, Berkeley, California 94720, USA}

\author{John Mark Kreikebaum}
\affiliation{Department of Physics, University of California, Berkeley, California 94720, USA}
\affiliation{Materials Science Division, Lawrence Berkeley National Laboratory, Berkeley, California 94720, USA}

\author{David I. Santiago}
\affiliation{Computational Research Division, Lawrence Berkeley National Laboratory, Berkeley, California 94720, USA}
\affiliation{Department of Physics, University of California, Berkeley, California 94720, USA}

\author{Irfan Siddiqi}
\affiliation{Computational Research Division, Lawrence Berkeley National Laboratory, Berkeley, California 94720, USA}
\affiliation{Department of Physics, University of California, Berkeley, California 94720, USA}
\affiliation{Materials Science Division, Lawrence Berkeley National Laboratory, Berkeley, California 94720, USA}

\date{\today}
             
\begin{abstract}
	The development of noisy intermediate-scale quantum (NISQ) devices has extended the scope of executable quantum circuits~\cite{Preskill18,Bharti21} with high-fidelity single- and two-qubit gates~\cite{Sung21,Kandala20,Brad21,Wei21}. Equipping NISQ devices with three-qubit gates will enable the realization of more complex quantum algorithms~\cite{Haner17,Gidney21,Figgatt17} and efficient quantum error correction protocols~\cite{Reed12,Paetznick13, Yoder16} with reduced circuit depth.
    Several three-qubit gates have been implemented for superconducting qubits~\cite{Fedorov12, Reed12,Hill21}, but their use in gate synthesis has been limited due to their low fidelity.
	Here, using fixed-frequency superconducting qubits, we demonstrate a high-fidelity $i$Toffoli gate based on two-qubit interactions, the so-called cross-resonance effect~\cite{Rigetti10,Chow11}.
	As with the Toffoli gate, this three-qubit gate can be used to perform universal quantum computation~\cite{Toffoli80,Shi03, Aharonov03}.
	The $i$Toffoli gate is implemented by simultaneously applying microwave pulses to a linear chain of three qubits, revealing a process fidelity as high as 98.26(2)\%. 
	Moreover, we numerically show that our gate scheme can produce additional three-qubit gates which provide more efficient gate synthesis than the Toffoli and $i$Toffoli gates.
	Our work not only brings a high-fidelity $i$Toffoli gate to current superconducting quantum processors but also opens a pathway for developing multi-qubit gates based on two-qubit interactions.
\end{abstract}
\maketitle

The three-qubit Toffoli gate is universal for reversible classical computation, enabling arbitrary Boolean operations over quantum registers~\cite{Toffoli80}. Moreover, together with the Hadamard gate, it forms a universal quantum gate set~\cite{Shi03, Aharonov03}. Appending the Toffoli gate to a gate set consisting of the CNOT and single-qubit gates is practically helpful for reducing the overhead of gate synthesis. For instance, both the overall gate count and the circuit depth can be reduced by a factor of $\mathcal{O}(\log{n})$ in Shor's algorithm, which factorizes $n$-bit integers~\cite{Haner17,Gidney21}. The oracle functions of Grover's search algorithm~\cite{Figgatt17} and quantum error correction protocols~\cite{Yoder16,Paetznick13,Reed12} can also be constructed with reduced resource costs. However, implementing a high-fidelity Toffoli gate is experimentally challenging; thus, the operation is usually realized by decomposing it into single- and two-qubit gates~\cite{Figgatt17,Figgatt19}. This decomposition requires at least five two-qubit gates for fully connected qubits~\cite{Barenco95,Yu13} and eight for nearest-neighbor connected qubits~\cite{Smith21}, considerably reducing its fidelity.

Native Toffoli gates have been implemented by exploiting specific features of each quantum computing platform, such as the common vibrational mode of ion strings~\cite{Monz09}, Rydberg blockade of neutral atoms~\cite{Levine19}, and electron spin resonance of quantum dots~\cite{Hendrickx21}. In superconducting circuits, the second or third excited states of frequency-tunable qubits have been employed~\cite{Fedorov12, Reed12,Hill21}, but the best fidelity was limited to $\sim$87\% as higher energy levels are generally more susceptible to decoherence and state leakage. Although a high-fidelity Toffoli gate has been  recently reported using all-to-all intermodal couplings of multimode superconducting circuits~\cite{Roy20}, the multimode architecture restricts the scalability of the gate. Additionally, there are proposals based on the degeneracy of excited-state manifolds~\cite{Khazali20}, Hamiltonian engineering~\cite{Zahedinejad15}, or strong Ising-type couplings~\cite{Rasmussen20}, which are challenging to implement experimentally.

Here, we demonstrate a high-fidelity $i$Toffoli gate by simultaneously driving three superconducting qubits at the same frequency. The simplicity of our scheme makes the gate calibration straightforward, and the gate can even be implemented on cloud-based quantum processors~\cite{IBM}. Similar to the Toffoli gate, the $i$Toffoli gate inverts a target qubit conditioned upon two control qubits but with a phase shift of $\pi/2$. This three-qubit gate can also be used for arbitrary Boolean operations~\cite{Toffoli80} and forms a universal quantum gate set with the Hadamard gate~\cite{Aharonov03}. To characterize our gate performance independently of state-preparation-and-measurement (SPAM) errors, we leverage cycle benchmarking~\cite{Erhard19} and measure a process fidelity of 98.26(2)\%. By analyzing the error budget, we find that the fidelity could be further improved using recently developed schemes to suppress undesired longitudinal interactions~\cite{Kandala20,Brad21,Wei21}. 
In addition, our numerical study shows that our gate scheme is capable of generating additional three-qubit gates that synthesize random unitary circuits with a lower circuit depth compared to the Toffoli and $i$Toffoli gates.

\begin{figure*}[t]
\centering
\includegraphics[width=1.97\columnwidth]{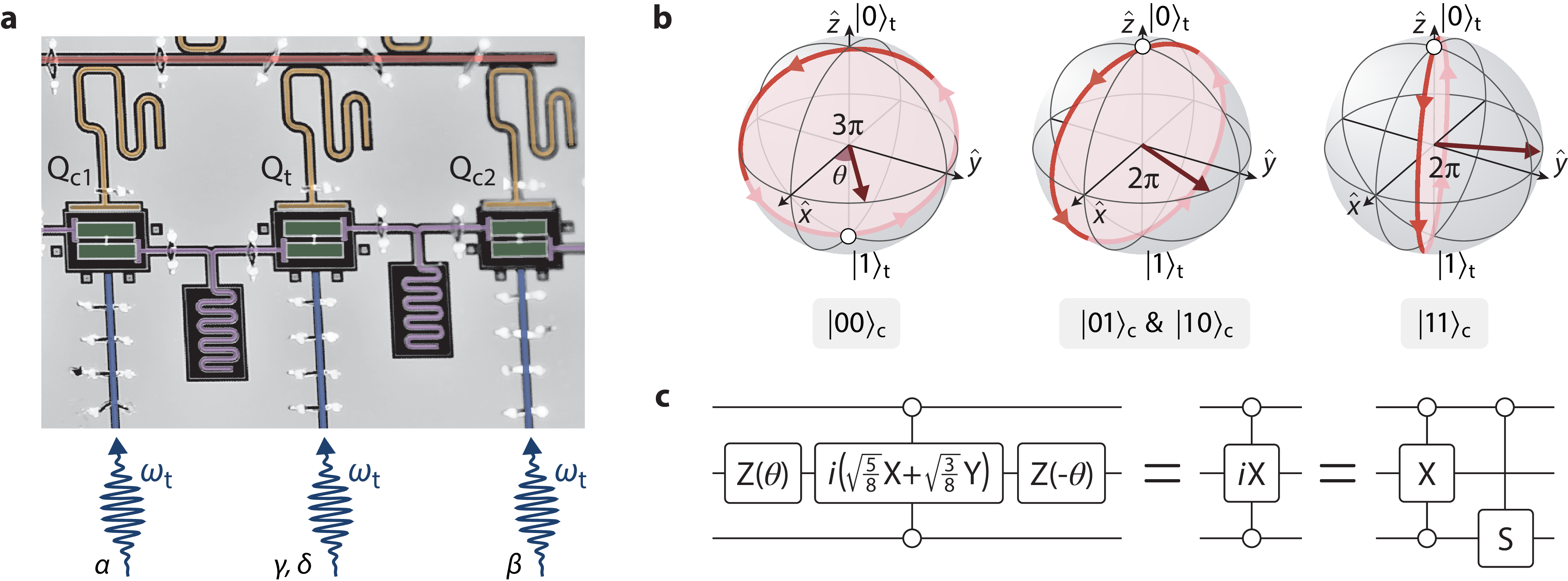}
\caption{\textbf{Experimental schematic and quantum circuit of the \textit{i}Toffoli gate.}
	\textbf{a}, False-colored micrograph of the quantum processor used in the experiment. The $i$Toffoli gate is performed on three transmon qubits (green), and the target qubit $\text{Q}_\text{t}$ is in between the control qubits $\text{Q}_\text{c1}$ and $\text{Q}_\text{c2}$. The qubits are capacitively coupled to their nearest-neighbors via coplanar stripline resonators (purple) and have independent microwave control lines (blue). To implement the $i$Toffoli gate, microwave pulses are simultaneously applied to all three qubits at the target qubit frequency $\omega_\text{t}$. After the gate operation, the qubit states are dispersively read out via coplanar waveguide readout resonators (yellow)~\cite{Wallraff04}. To reduce the footprint, the readout resonators are coupled to a dual-purpose common bus (red) which allows multiplexed readout and protects the qubits from Purcell decay.
	\textbf{b}, The conditional state evolution of $\text{Q}_\text{t}$ during the $i$Toffoli operation. The target state initially in $|0\rangle_\text{t}$ rotates on the Bloch sphere depending on the control state of $\text{Q}_\text{c1}$ and $\text{Q}_\text{c2}$. The brown arrow and white dot represent the rotation axis and the final target state after the $i$Toffoli operation, respectively. See details in main text.
	\textbf{c}, Circuit diagram to realize the $i$Toffoli gate. The rotation axis in $\textbf{b}$ for $|00\rangle_\text{c}$ is effectively transformed onto the $x$-axis by sandwiching the gate operation in between virtual Z gates. The $i$Toffoli and the Toffoli gates differ only by a controlled-phase gate between $\text{Q}_\text{c1}$ and $\text{Q}_\text{c2}$, which imparts a $\pi/2$ phase shift to $|00\rangle_{\text{c}}$. 
	}
\label{fig1} 
\end{figure*}

\vspace{0.7em}
\noindent\textbf{Schematic of the \textit{i}Toffoli gate}\\
The $i$Toffoli gate is performed on the three fixed-frequency transmons~\cite{Koch07} shown in Fig.~\ref{fig1}\textbf{a}. We use the left and right transmons as the control qubits and the center one as the target qubit, labeled $\text{Q}_\text{c1}$, $\text{Q}_\text{c2}$, and $\text{Q}_\text{t}$ accordingly. The device parameters and the experimental setup are presented in Methods and Supplementary Note 1, respectively. The $i$Toffoli gate is implemented by simultaneously applying microwave pulses to the three qubits at the target qubit frequency $\omega_\text{t}$. The microwave tones on the control qubits $\text{Q}_\text{c1,c2}$ induce the cross-resonance effect~\cite{Rigetti10,Chow11} between $\text{Q}_{\text{c1,c2}}$ and $\text{Q}_\text{t}$ that manifests the interaction Hamiltonians $H_\text{CR}\propto \text{Z}_{\text{c1,c2}} \text{X}_\text{t}$, while the tone on the target qubit results in $\text{X}_\text{t}$ and $\text{Y}_\text{t}$ rotations. $\text{X}_k$, $\text{Y}_k$, and $\text{Z}_k$ represent the Pauli operators for qubit $\text{Q}_k$. Taken together, the Hamiltonian of the driven three-qubit system is,
\begin{equation}
    H =\alpha\ \! \text{Z}_{\text{c1}}\text{X}_\text{t}+\beta\ \!\text{X}_\text{t}\text{Z}_{\text{c2}} 
    +\gamma\ \! \text{X}_\text{t}+\delta\ \!\text{Y}_\text{t},
    \label{HTOF}
\end{equation}
where $\alpha$ and $\beta$ depend linearly on the amplitude of the control drives, and $\gamma$ and $\delta$ are determined by the target drive's amplitude and phase. Here we have omitted $\text{Z}_{\text{c1}}$ and $\text{Z}_{\text{c2}}$ terms, caused by the off-resonant control drives, as they can be removed at no cost using virtual Z gates~\cite{McKay17}.

A Toffoli-like gate is implemented if $\text{Q}_\text{t}$ undergoes a $\pi$ rotation about the $x$-axis on the Bloch sphere for a certain control state and remains invariant otherwise. To explore the conditional evolution of $\text{Q}_\text{t}$, we reduce the Hamiltonian $H$ in Eq.~(\ref{HTOF}) into the effective target Hamiltonian $H^{kl}_{\text{t}}\equiv~\!\! _{\text{c}}\langle kl|H|kl\rangle_{\text{c}}$ depending on control state $|kl\rangle_\text{c}$:
\begin{eqnarray}
    H^{00}_{\text{t}}&=&(\alpha+\beta+\gamma)\text{X}_\text{t}+\delta\ \! \text{Y}_\text{t},\nonumber\\
    H^{01}_{\text{t}}&=&(\alpha-\beta+\gamma)\text{X}_\text{t}+\delta\ \! \text{Y}_\text{t},\nonumber\\
    H^{10}_{\text{t}}&=&(-\alpha+\beta+\gamma)\text{X}_\text{t}+\delta\ \! \text{Y}_\text{t},\nonumber\\
    H^{11}_{\text{t}}&=&(-\alpha-\beta+\gamma)\text{X}_\text{t}+\delta\ \! \text{Y}_\text{t}.
    \label{H_target}
\end{eqnarray} 
Although the conditional inverting operation can be realized for any control state with these Hamiltonians, in this work, we demonstrate the conditional operation for $|00\rangle_\text{c}$. Specifically, we consider the scheme in which $\text{Q}_\text{t}$ rotates by $3\pi$ when the control state is $|00\rangle_\text{c}$ and by $2\pi$ otherwise. This can be accomplished by first imposing $\alpha = \beta =\gamma$ to make the conditional Rabi frequencies of $\text{Q}_\text{t}$ the same except for $|00\rangle_\text{c}$: $\Omega^{00}_\text{t} =\sqrt{9\alpha^2 + \delta^2}$ for $|00\rangle_\text{c}$ and $\Omega^{\text{oth}}_\text{t}=\sqrt{\alpha^2 + \delta^2}$ for the other control states. Then, we can make $\Omega^{00}_\text{t} = 1.5\Omega^{\text{oth}}_\text{t}$ with $\delta = \sqrt{27/5}\alpha$, and a Toffoli-like gate is realized by setting the gate duration to $\tau =2\pi/\Omega^{\text{oth}}_\text{t}$.

Figure~\ref{fig1}\textbf{b} shows the conditional evolution of $\text{Q}_\text{t}$ from the initial state $|0\rangle_\text{t}$. Due to the Pancharatnam-Berry geometric phase~\cite{Jordan88,Cho19}, the target qubit accumulates a $3\pi/2$ phase for control state $|00\rangle_\text{c}$ and a $\pi$ phase for the other control states, making the operation the $i$Toffoli gate rather than the conventional Toffoli gate. Although the rotation axis for $|00\rangle_\text{c}$ is not aligned along the $x$-axis, this axis can effectively be transformed onto the $x$-axis by sandwiching the three-qubit operation with virtual Z gates on the target qubit. The circuit diagram of the $i$Toffoli operation is shown in Fig.~\ref{fig1}\textbf{c}. The $i$Toffoli gate differs from the Toffoli gate only by a controlled-phase gate~\cite{Selinger13}. 

\begin{figure}[t!]
    \centering
    \includegraphics[width=1\columnwidth]{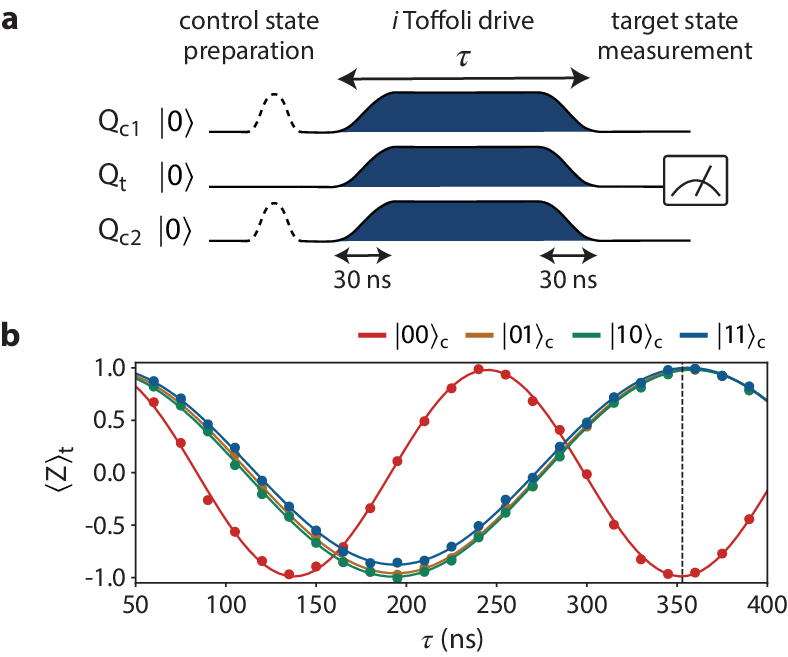}
    \caption{\textbf{Pulse sequence and conditional Rabi oscillations of the target qubit.} \textbf{a}, For the $i$Toffoli gate, the three qubits are simultaneously driven using flat-top cosine-ramp pulses (blue) at the transition frequency of $\text{Q}_\text{t}$. To investigate the conditional evolution of $\text{Q}_\text{t}$, the control state of $\text{Q}_\text{c1}$ and $\text{Q}_\text{c2}$ is initialized in $|00\rangle_\text{c}$, $|01\rangle_\text{c}$, $|10\rangle_\text{c}$, and $|11\rangle_\text{c}$, and then the expectation value of $\text{Q}_\text{t}$ is measured in the Pauli Z basis. The gate calibration procedure is detailed in Methods. \textbf{b}, After the gate calibration, the conditional Rabi oscillations of $\text{Q}_\text{t}$ initially in $|0\rangle_\text{t}$ are measured as functions of the pulse duration $\tau$. The gate duration of the $i$Toffoli gate (dashed line) is determined by fitting the oscillations.}
    \label{fig3} 
\end{figure}

\vspace{1em}
\noindent\textbf{Gate operation and geometric phase error}\\
We  now  turn  to  the experimental implementation of the $i$Toffoli gate. For the gate operation, as shown in Fig~\ref{fig3}\textbf{a}, we use microwave pulses with a flat-top envelope and 30-ns cosine ramps. The pulses are applied at the target qubit frequency $\omega_\text{t}$ for pulse duration $\tau$. The gate calibration is conducted by tuning the amplitude and phase of each pulse to make the conditional Rabi frequencies $\Omega_\text{t}^{00}=1.5\Omega_\text{t}^{01}=1.5\Omega_\text{t}^{10}=1.5\Omega_\text{t}^{11}$. The calibration procedure is detailed in Methods. Figure~\ref{fig3}\textbf{b} shows the experimental results of the conditional Rabi oscillations starting from $|0\rangle_\text{t}$, and the $i$Toffoli gate is implemented at $\tau=353~\mathrm{ns}$.

The oscillation amplitudes of less than 1 in Fig.~\ref{fig3}\textbf{b} indicate that the rotation axis is not in the $xy$-plane. The axial tilt is mainly due to parasitic longitudinal (ZZ) interactions between nearest-neighbor qubits~\cite{Tripathi19,Magesan20}. These undesired interactions add a conditional $\Delta^{kl} \text{Z}_\text{t}$ term to the effective target Hamiltonians in Eq.~(\ref{H_target}) for each control state $|kl\rangle_\text{c}$. This is detrimental to the $i$Toffoli gate as a non-zero $\Delta^{kl}$ induces a different geometric phase (GP) for $|0\rangle_\text{t}$ and $|1\rangle_\text{t}$. Note that the GP is given by $\frac{1}{2}\int_{\mathcal{C}}(1-\cos{\theta})d\phi$ for a trajectory $\mathcal{C}$ on the Bloch sphere, where $\theta$ is the angle between the state vector and the rotation axis and $\phi$ denotes the rotation angle~\cite{Jordan88,Cho19}. Considering the rotation axis tilted by undesired interactions, $\cos{\theta}$ is $\Delta^{kl}/\Omega^{kl}_\text{t}$ for $|0\rangle_\text{t}$ and $-\Delta^{kl}/\Omega^{kl}_\text{t}$ for $|1\rangle_\text{t}$. Thus, in the absence of qubit decoherence, the $i$Toffoli operation $U_\text{GP}$ maps the computational basis states as follows:
\begin{eqnarray}
|kl\rangle_\text{c}|0\rangle_\text{t} &\rightarrow& e^{i\pi(1-\Delta^{kl}/\Omega^{kl}_\text{t})}|kl\rangle_\text{c}|0\rangle_\text{t},\nonumber\\
|kl\rangle_\text{c}|1\rangle_\text{t} &\rightarrow& e^{i\pi(1+\Delta^{kl}/\Omega^{kl}_\text{t})}|kl\rangle_\text{c}|1\rangle_\text{t},
\label{GP}
\end{eqnarray}
where $|kl\rangle_\text{c}\in\{|01\rangle_\text{c},|10\rangle_\text{c},|11\rangle_\text{c}\}$. In the case of $|00\rangle_\text{c}$, $\Delta^{00}$ is presumed to be zero so that the target qubit's computational states are perfectly inverted with a phase shift of $3\pi/2$. The condition $\Delta^{00}=0$ is satisfied if the drive tones are at the target qubit frequency $\omega_\text{t}$. We estimate the GP error in $U_\text{GP}$ to be $1-F(U_\text{GP},U_\text{ideal})=0.31\%$ by obtaining $\Delta^{kl}/\Omega^{kl}_\text{t}$ from the oscillation amplitudes in Fig.~\ref{fig3}\textbf{b}. Here the entanglement fidelity is adopted as the measure of the process fidelity, $F(V,U)=|\text{Tr}[V^{\dagger}U]/d|^2$ where $d=8$ is the Hilbert space dimension~\cite{Nielsen02}.


\begin{figure}[t]
\centering
\includegraphics[width=1\columnwidth]{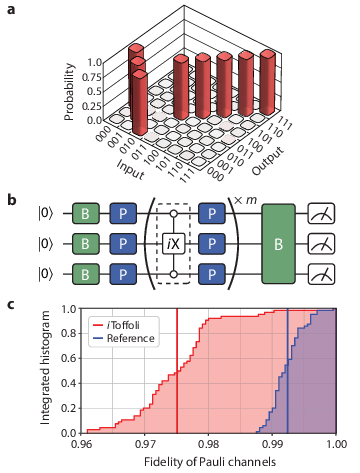}
\caption{\textbf{Benchmark of a 353-ns \textit{i}Toffoli gate.} \textbf{a}, Truth table of the $i$Toffoli gate. The input and output states denote the states of $\text{Q}_\text{c1}$, $\text{Q}_\text{t}$, and $\text{Q}_\text{c2}$ in order. The red bars and the solid borders represent the experimental and ideal results, respectively. \textbf{b}, Circuit schematic of cycle benchmarking (CB). The $i$Toffoli gate is twirled with random Pauli gates (blue) to tailor the gate errors to stochastic Pauli error channels. The initial state and measurement basis (green) are chosen to pick up an error of a certain Pauli channel using True-Q~\cite{Beale20}. The Pauli-twirling error is separately estimated by performing CB without the $i$Toffoli gate as a reference. \textbf{c}, The cumulative fidelity histograms of all three-qubit Pauli channels for the Pauli-twirled $i$Toffoli gate and the reference, and the solid vertical lines represent the average fidelities of $97.51(2)\%$ and $99.24(1)\%$. The process fidelity of the $i$Toffoli gate is given as $98.26(2)\%$ by deducting the Pauli-twirling error.} 
\label{fig4} 
\end{figure}

\vspace{1em}
\noindent\textbf{Gate benchmark and error budget analysis}\\
As an initial characterization of the $i$Toffoli gate, the classical truth table is obtained for the computational states after correcting the readout errors based on the readout fidelities~\cite{Nachman20} (see Fig.~\ref{fig4}\textbf{a}). This table shows that the target state is inverted only when the control state is $|00\rangle_\text{c}$. The process fidelity of the truth table $|U_\text{exp}|$, excluding the GP error, is given as $F(|U_\text{exp}|,|U_\text{ideal}|)=97.39(8)\%$ with respect to the ideal operation $|U_\text{ideal}|$. The uncertainty is obtained by performing 1000 Monte Carlo simulation runs taking into account statistical errors. The population leakage to the second excited state is measured to be less than 0.1\% for all the qubits.

The fidelity measured above is subject to state-preparation-and-measurement (SPAM) errors. To extract a SPAM-error-free fidelity estimate of a quantum operation, interleaved randomized benchmarking (IRB) is typically used~\cite{Magesan11,Magesan12,Morvan21}. However, IRB is impractical for the $i$Toffoli gate due to the complexity of three-qubit Clifford twirling~\cite{McKay19}. As an alternative, we perform cycle benchmarking (CB) which relies on Pauli twirling~\cite{Erhard19}. The CB protocol has better scalability than IRB as the complexity of twirling operation is reduced. The circuit schematic of the CB protocol is depicted in Fig.~\ref{fig4}\textbf{b}. The Pauli twirling of the $i$Toffoli gate tailors the gate errors to stochastic Pauli error channels, and the initial state and measurement basis are chosen to pick up a Pauli channel using True-Q~\cite{Beale20}.

We run CB for all three-qubit Pauli channels with depth $m\in\{2,4,16,32\}$ and 30 samples for each $m$. In order to decouple SPAM errors from fidelity estimate, the fidelity $p_k$ of each Pauli channel $k$ is extracted by fitting the fidelity decay to an exponential model $A p_k^m$. However, since the fidelity still includes the error from Pauli-twirling operations, the Pauli-twirling error needs to be separately estimated by performing CB without the $i$Toffoli gate as a reference. Figure~\ref{fig4}\textbf{c} shows the cumulative fidelity histograms of all three-qubit Pauli channels for the Pauli-twirled $i$Toffoli gate $p_k^{(\text{itof})}$ and the reference $p_k^{(\text{ref})}$. The fidelity of each Pauli channel is exhibited in Supplementary Note 2. From the results, the process fidelity of the $i$Toffoli gate is estimated to be $98.26(2)\%$ by averaging $p^{(\text{itof})}_k/p^{(\text{ref})}_k$ over all the Pauli channels~\cite{Erhard19}. The uncertainty represents the fitting errors of fidelity decays. 


\begin{figure}[t]
    \centering
    \includegraphics[width=0.95\columnwidth]{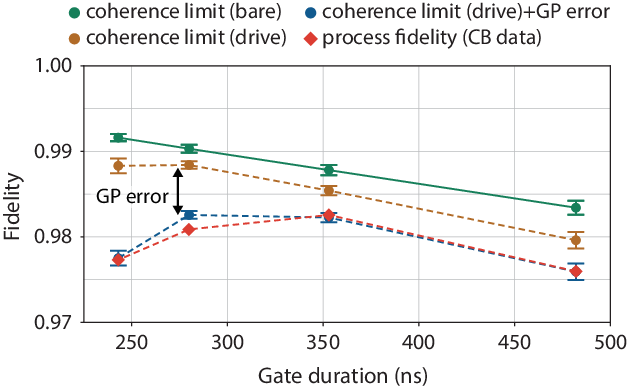}
    \caption{\textbf{Error budget of \textit{i}Toffoli gates having different gate durations.} The process fidelities of the \textit{i}Toffoli gates (red) are experimentally estimated via cycle benchmarking (CB) after calibrating the gates using different drive amplitudes. The solid green line shows the coherence limit based on the measured $T_1$ and $T_2^\text{echo}$. The coherence times are reduced under the $i$Toffoli drive, and the brown dots show this reduced coherence limit at each gate duration. The sum of the geometric phase (GP) error estimated from Eq.~(\ref{GP}) and the coherence-limited infidelity under drive (blue) agrees well with the CB data (red). The errors of CB data are obtained from the fitting errors and are smaller than the marker size, and the other errors are estimated from the statistical fluctuations of the coherence times.}
    \label{fig5} 
\end{figure}

\begin{figure}[h]
    \centering
    \includegraphics[width=1\columnwidth]{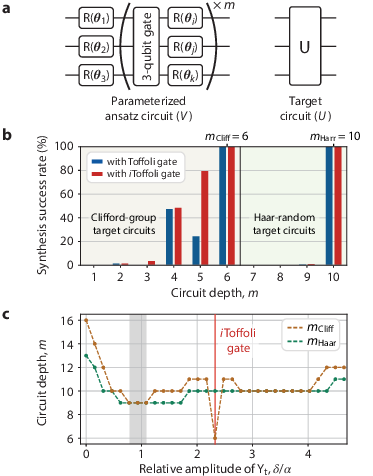}
    \caption{\textbf{Gate synthesis of three-qubit circuits.} \textbf{a}, An ansatz circuit $V$ is constructed by alternating parameterized SU(2) gates~\cite{McKay17} and a fixed three-qubit gate with depth $m$. For gate synthesis of a target circuit $U$, the ansatz parameters are optimized by minimizing the distance between $V$ and $U$, or their infidelity $1-F(V,U)$, via True-Q~\cite{Beale20}. \textbf{b}, We synthesize 1000 Clifford-group target circuits (1$\le\!m\!\le$6) or 1000 Haar-random target circuits (7$\le\!m\!\le$10) at each~$m$. The target circuits are sampled using QuTiP~\cite{qutip}. The bar graph shows success rates of the gate syntheses when adopting either the Toffoli (blue) or the $i$Toffoli (red) gate as the ansatz three-qubit gate. Here we define the synthesis success as reaching the numerical precision after optimizing synthesis infidelity. The threshold depths $m_\text{Cliff}$ and $m_\text{Haar}$ represent the circuit depths providing 100\% synthesis success rate. 
    \textbf{c}, We estimate threshold depths $m_\text{Cliff}$ (brown) and $m_\text{Haar}$ (green) of our gate scheme by varying the amplitudes of $\text{Y}_\text{t}$ in Eq.~(\ref{HTOF}). The other Hamiltonian amplitudes and the gate duration are fixed to the same values used for implementing the $i$Toffoli gate. Note that, as finite Haar-random unitary sampling cannot cover all three-qubit circuits including Clifford-group circuits, $m_\text{Haar}$ could be less than $m_\text{Cliff}$. The greater of $m_\text{Cliff}$ and $m_\text{Haar}$ corresponds to the threshold depth for synthesizing arbitrary unitary circuits. The shaded region represents the amplitude range of $\text{Y}_\text{t}$ providing the lowest threshold depth, and the red solid line shows $\delta/\alpha=\sqrt{27/5}$ for the $i$Toffoli gate.}
    \label{fig2} 
\end{figure}

Additionally, to get insights into factors limiting improvement in process fidelity, we perform CB for $i$Toffoli gates having different gate durations and analyze the error budget. As the error sources, the geometric phase (GP) error and decoherence of the qubits are considered. The GP error is estimated according to Eq.~(\ref{GP}), and the coherence-limited infidelity~\cite{Dawkins20} is evaluated using the relaxation ($T_1$) and the Hahn-echo dephasing ($T_2^\text{echo}$) times of each qubit and the gate duration. See Methods for the calculation of coherence-limited infidelity. We experimentally observed that the $T_2^\text{echo}$ of $\text{Q}_\text{c1}$ and $\text{Q}_\text{c2}$ are reduced under the $i$Toffoli drive relative to the bare coherence times. Accordingly, we calculated the coherence limits using both the bare coherence times and the coherence times under drive. The measurement sequences and data of the coherence times are presented in Supplementary Note 3. Figure~\ref{fig5} shows that the process fidelities obtained from CB are well explained by the estimated error budget. The results reveal that undesired longitudinal interactions, or the GP errors, increase with decreasing gate duration, limiting further reduction in gate duration and improvement in process fidelity.


\vspace{1em}
\noindent\textbf{Gate synthesis of three-qubit circuits}\\
Finally, we numerically explore gate synthesis of three-qubit circuits to demonstrate the utility of the $i$Toffoli gate. To this end,  we consider an ansatz circuit $V$ depicted in Fig.~\ref{fig2}\textbf{a} with depth $m$. The gate synthesis of a target circuit $U$ is conducted by optimizing the ansatz parameters to minimize the distance between $V$ and $U$, or their infidelity $1-F(V,U)$. The simulation results in Fig.~\ref{fig2}\textbf{b} show that both Toffoli and $i$Toffoli gates can be used to synthesize arbitrary three-qubit Clifford circuits at $m_\text{Cliff}=6$ and Haar-random unitary circuits at $m_\text{Haar}=10$. Additionally, the results indicate that the ansatz circuit with the $i$Toffoli gate can synthesize more various target circuits at lower depths than the Toffoli gate. 

Furthermore, we estimate threshold depths $m_\text{Cliff}$ and $m_\text{Haar}$ of our gate scheme by varying $\text{Y}_\text{t}$'s amplitude $\delta$ in the Hamiltonian of Eq.~(\ref{HTOF}). For this simulation, the other Hamiltonian amplitudes and the gate duration are fixed to the same values used for implementing the $i$Toffoli gate. As $\text{Y}_\text{t}$ is the only non-commuting term in the Hamiltonian, it plays a key role in preventing the Hamiltonian operation from being a simple combination of single- and two-qubit gates. Interestingly, the results in Fig.~\ref{fig2}\textbf{c} reveal that, over a range of $\delta$ values, our gate scheme can produce three-qubit gates which synthesize random unitary circuits at $m_\text{Cliff}=m_\text{Haar}=9$. That is, while the structured Toffoli and $i$Toffoli gates have advantages for Clifford circuits, these alternative three-qubit gates could be beneficial for implementing random unitary circuits.

\vspace{1em}
\noindent\textbf{Discussion and outlook}\\
We have demonstrated a single-step $i$Toffoli gate scheme based on the cross-resonance (CR) effect with fixed-frequency transmon qubits. 
As the CR effect can readily be implemented between capacitively coupled superconducting qubits~\cite{Rigetti10,Chow11}, our scheme is applicable to a number of superconducting architectures including cloud-based quantum processors~\cite{IBM}. In our gate scheme, the non-commuting term $\text{Y}_\text{t}$ in Eq.~(\ref{HTOF}) plays a key role in preventing the full Hamiltonian operation from being decomposable into simple consecutive operations. By varying the amplitude of the $\text{Y}_\text{t}$ term, we have numerically found that this gate scheme can produce additional three-qubit gates providing more efficient gate synthesis than both the Toffoli and $i$Toffoli gates. We hope our work will trigger further studies to develop diverse multi-qubit gates derived from two-qubit interactions involving non-commuting Hamiltonian operators.

To characterize a 353-ns $i$Toffoli gate without state-preparation-and-measurement (SPAM) errors, we have leveraged cycle benchmarking and measured a process fidelity of 98.26(2)\%. The infidelity is primarily due to decoherence of the qubits and undesired longitudinal (ZZ) interactions. By analyzing the error budget, we have found that such undesired interactions limit a further improvement in process fidelity. Recently, the multi-path coupling scheme~\cite{Kandala20} and the microwave-induced ZZ tuning scheme~\cite{Brad21,Wei21} have been demonstrated to suppress ZZ interactions and improve the fidelity of two-qubit CR gates. These schemes will directly benefit the fidelity of our CR-based $i$Toffoli gate as well. We expect that our high-fidelity three-qubit gate will open a pathway for executing complex quantum circuits, such as Shor's and Grover's algorithms~\cite{Haner17,Gidney21,Figgatt17}, and for exploring efficient quantum error correction protocols~\cite{Reed12,Paetznick13, Yoder16}.

\vspace{1em}
\secNATend{Methods}
\noindent\textbf{Device parameters}\\
The transition frequencies and anharmonicities of transmon qubits $\{\text{Q}_\text{c1}, \text{Q}_\text{t}, \text{Q}_\text{c2} \}$ are measured as $\{ 5.254, 5.331, 5.491\}$ GHz and $\{ -277.1, -272.1, -271.8 \}$ MHz, respectively, using Ramsey spectroscopy. Their coherence times are estimated through the measurements of the excited state lifetime and the Hahn echo decay: $T_1 = \{70(7), 61(3), 57(9) \}~\mathrm{\mu s}$ and $T_2^\text{echo} = \{ 60(5), 73(9), 66(9) \}~\mathrm{\mu s}$. To obtain the static longitudinal (ZZ) interaction strength between pairs of qubits, conditional Ramsey measurements are conducted, yielding $\zeta_{ZZ}/2\pi = 96~\mathrm{kHz}$ for $\{\text{Q}_\text{c1}, \text{Q}_\text{t} \}$ and $\zeta_{ZZ}/2\pi = 171~\mathrm{kHz}$ for $\{ \text{Q}_\text{t},\text{Q}_\text{c2}\}$. The readout frequencies are measured as  $\{6.564,6.623,6.679\}$ GHz, and the readout fidelities are obtained as $P(0|0)=\{0.998,0.997,0.996\}$ and $P(1|1)=\{0.948,0.982,0.981\}$, where $P(x|y)$ is the~probability that the qubit in state $y$ is measured to be in state~$x$.

\vspace{1em}
\noindent\textbf{Calibration procedure}\\ 
As shown in Fig~\ref{fig3}\textbf{a}, individual microwave pulses are applied to all three qubits. Using the effective amplitude $a_k$ and phase $\phi_k$ of each pulse, we can recast the three-qubit Hamiltonian in Eq.~(\ref{HTOF}) into,
\begin{eqnarray}
    H =&&a_{\text{c1}}\cos{\phi_{\text{c1}}}\ \!\text{Z}_{\text{c1}}\text{X}_\text{t}+a_{\text{c1}}\sin{\phi_{\text{c1}}}\ \!\text{Z}_{\text{c1}}\text{Y}_\text{t}\nonumber\\
    &+&a_{\text{c2}}\cos{\phi_{\text{c2}}}\ \!\text{X}_\text{t}\text{Z}_{\text{c2}}+a_{\text{c2}}\sin{\phi_{\text{c2}}}\ \!\text{Y}_\text{t}\text{Z}_{\text{c2}}\nonumber\\
    &+&a_{\text{t}}\cos{\phi_{\text{t}}}\ \!\text{X}_\text{t}+a_{\text{t}}\sin{\phi_{\text{t}}}\ \!\text{Y}_\text{t}.
    \label{Hdrive}
\end{eqnarray}
 In order to implement the $i$Toffoli gate, the pulse parameters need to be calibrated to satisfy $\alpha=\beta=\gamma=\sqrt{5/27}\ \!\delta$ in terms of Eq.~(\ref{HTOF}). Whether the condition is satisfied can be checked by measuring the conditional Rabi frequencies $\Omega^{kl}_\text{t}$ of $\text{Q}_\text{t}$ initially in $|0\rangle_\text{t}$. The calibration procedure of the $i$Toffoli gate is as follows:
\begin{enumerate}[itemsep=0ex]
\item  \textbf{Set} $\boldsymbol{\alpha}$: Fix $a_\text{c1}$ and $\phi_\text{c1}$ of the first control drive. The gate duration will be inversely proportional to the amplitude $a_\text{c1}$.

\item \textbf{Set} $\boldsymbol{\beta=\alpha}$: Tune $a_\text{c2}$ and $\phi_\text{c2}$ of the second control drive to make $\Omega_\text{t}^{01}=\Omega_\text{t}^{10}$.

\item \textbf{Remove ZY}: Sweep $\phi_{c1}$ and $\phi_{c2}$ simultaneously with the same offset to remove $\text{Z}_{\text{c1}}\text{Y}_{\text{t}}$ and $\text{Y}_{\text{t}}\text{Z}_{\text{c2}}$. The amplitudes are estimated by performing Hamilton tomography~\cite{Sheldon16}. Note that the simultaneous phase sweep will not change the condition of $\Omega_\text{t}^{01}=\Omega_\text{t}^{10}$.

\item \textbf{Set} $\boldsymbol{\gamma=\alpha}$: Tune $a_\text{t}$ of the target drive with $\phi_\text{t}=0$, for $\text{X}_\text{t}$, to achieve $\Omega^{11}_\text{t}=\Omega^{01}_\text{t}$.

\item \textbf{Set} $\boldsymbol{\delta=\sqrt{27/5} \alpha}$: Add additional target drive for $\text{Y}_\text{t}$ with $\phi_\text{t}=\pi/2$ and tune the amplitude to make  $\Omega_\text{t}^{00} = 1.5\Omega_\text{t}^{01}$.
\end{enumerate}
If needed, the calibration procedure is repeated until $\Omega_\text{t}^{00} = 1.5\Omega^{01}_\text{t} = 1.5\Omega^{10}_\text{t} = 1.5\Omega^{11}_\text{t}$ is fulfilled. Even if there are undesired longitudinal interactions, the calibration procedure can be conducted in the same manner. Finally, the gate duration is determined so that $\text{Q}_\text{t}$ rotates by 3$\pi$ for control state $|00\rangle_\text{c}$.

\vspace{1em}
\noindent\textbf{Coherence-limited process fidelity}\\
A Pauli transfer matrix (PTM) of a Pauli-twirled single-qubit decoherence channel is written as~\cite{Dawkins20},
\begin{equation}
\mathcal{E} = 
        \begin{pmatrix}
        1 & 0 & 0 & 0 \\
        0 & \sqrt{1-\gamma_1}\sqrt{1-\gamma_2} & 0 & 0 \\
        0 & 0 & \sqrt{1-\gamma_1}\sqrt{1-\gamma_2} & 0 \\
        0 & 0 & 0 & 1-\gamma_1 
        \end{pmatrix},
\end{equation}
where $\gamma_1 = 1-e^{-\varGamma_1 \tau}$ and $\gamma_2 = 1-e^{-2\varGamma_2 \tau}$ represent the degree of relaxation and dephasing for gate duration $\tau$, respectively. The relaxation and dephasing rates of a qubit can be experimentally estimated by measuring the coherence times: $\varGamma_1=1/T_1$ and $\varGamma_2=1/T_2^{\text{echo}}-1/2T_1$. In the absence of crosstalk errors, the PTM of a Pauli-twirled 3-qubit channel is simply given by the tensor product $\mathcal{E}^{\otimes 3}$, and the coherence-limited process fidelity is calculated as,
\begin{eqnarray}
F_\text{cl}&=&\frac{1}{d^2}\text{Tr}[\mathcal{E}^{\otimes 3}]\\
&=&\frac{1}{d^2}\prod_{q=1}^3 \left(2 + 2\sqrt{1-\gamma^{(q)}_1}\sqrt{1-\gamma^{(q)}_2}-\gamma^{(q)}_1 \right), \nonumber
\end{eqnarray}
where $\gamma^{(q)}_1$ and $\gamma^{(q)}_2$ denote the decoherence of qubit $q$, and $d=8$ is the Hilbert space dimension.

\vspace{1em}
\secNATend{Acknowledgements}
    The authors gratefully acknowledge the conversations and insights of J. Wallman and I. Hincks.
    This work was supported by the Quantum Testbed Program of the Advanced Scientific Computing Research Division, Office of Science of the U.S. Department of Energy under Contract No. DE-AC02-05CH11231.

\end{document}


\section{Supplementary note 1 - Experimental setup}
\begin{figure}[h]
    \centering
    \includegraphics[width=0.79\columnwidth]{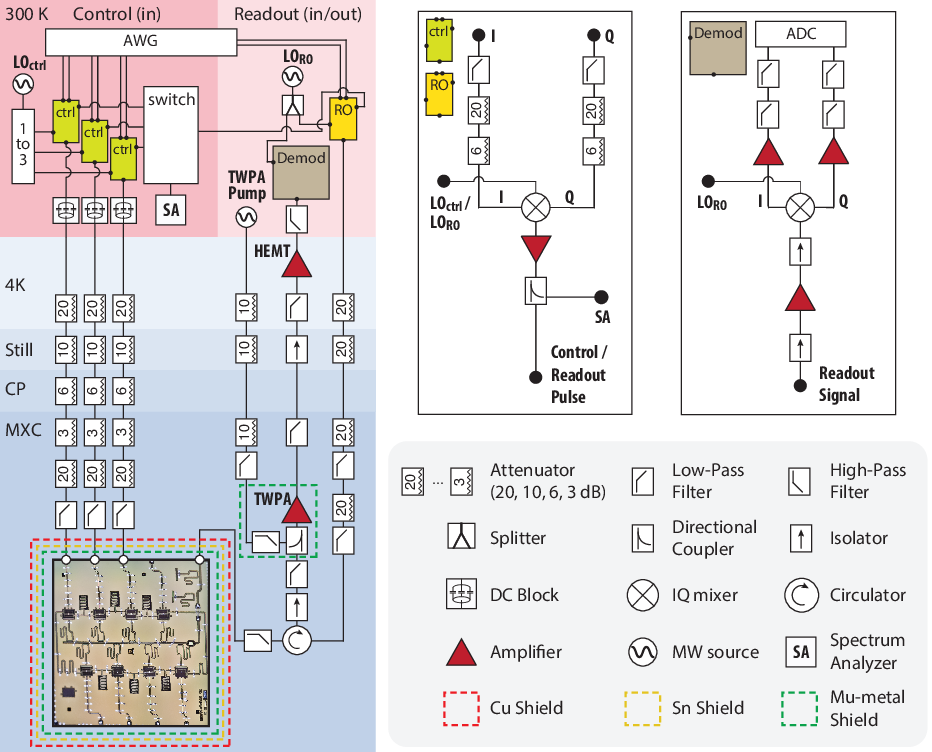}
    \caption{Experimental setup.}
    \label{fig:figS1}
\end{figure}
 A detailed experimental schematic is shown in Fig.~S\ref{fig:figS1}. We implement the $i$Toffoli gate on the 8-qubit superconducting quantum processor used in Ref.~\cite{Hashim20}. Specifically, in this work, we employ the top-left three transmon qubits in the processor image. The processor is operated in a BlueFors XLD dilution refrigerator with a base temperature of 10 mK and magnetically shielded with Copper (Cu), Tin (Sn), and Mu-metal cylinders. 
 
Qubit control and readout IF pulses are generated using a Keysight PXI Arbitrary Waveform Generator (AWG) at a sample rate of 1 GS/s. The IF pulses are then upconverted using a 5.415 GHz tone for control pulses and a 6.83 GHz tone for readout pulses via IQ mixers. Two Keysight MXG N5183B separately generate these local oscillator (LO) tones. To suppress thermal photons transmitted from higher-temperature stages, RF attenuators are installed at each stage and K\&L low-pass filters are additionally used at the base stage. 

After the readout pulses are dispersively coupled with qubit states, the readout signals are amplified by a traveling wave parametric amplifier (TWPA) at 10 mK, a high-electron mobility transistor amplifier (HEMT) at 4K, and room temperature amplifiers. The TWPA is pumped by a 16.5 dBm, 7.42 GHz tone (Hittite HMC M2100). To analyze the readout outcomes, the amplified readout signals are downconverted to IF signals with a 6.83 GHz LO tone and digitized by an Alazar ADC at a sample rate of 1 GS/s.

\section{Supplementary note 2 - Cycle Benchmarking and Pauli Transfer Matrix}
\begin{figure*}[h]
    \centering
     \includegraphics[width=1\columnwidth]{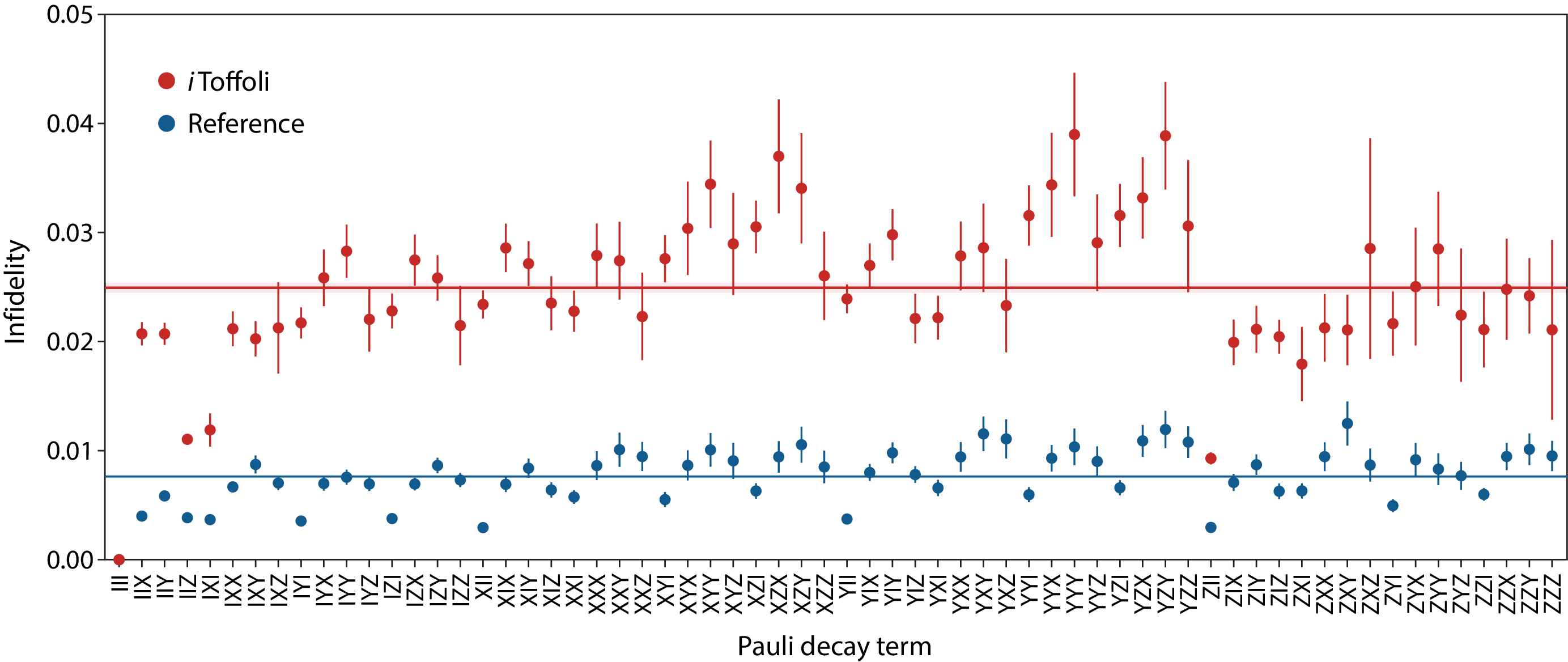}
    \caption{Detailed cycle benchmarking results of a 353-ns $i$Toffoli gate and the reference. The uncertainty of each data is obtained from the fitting error. The horizontal solid lines represent the averaged infidelities of the Pauli-twirled $i$Toffoli gate and the reference.}
    \label{fig:figS4}
\end{figure*}

We characterize the quantum operation of a 353-ns $i$Toffoli gate using two methods. First, as described in the main text, we run CB for all three-qubit Pauli channels with depth $m\in\{2,4,16,32\}$ and 30 samples for each $m$. The fidelity $p_k$ of each Pauli channel $k$ is extracted by fitting the fidelity decay to the exponential model $A p_k^m$. This allows decoupling the fidelity from state-preparation-and-measurement (SPAM) errors. The Pauli fidelities of the Pauli-twirled $i$Toffoli gate $p_k^{(\text{itof})}$ and the reference $p_k^{(\text{ref})}$ are shown in Figure~S\ref{fig:figS4}. The reference fidelities $p_k^{(\text{ref})}$ represent the errors from Pauli-twirling operations. By deducting the Pauli-twirling errors, the SPAM-error-free Pauli fidelities of the $i$Toffoli gate can be obtained as $p_k^{(\text{itof})}/p_k^{(\text{ref})}$. We estimate process fidelity by averaging Pauli fidelities over all three-qubit Pauli channels~\cite{Nielsen02}. Consequently, the process fidelities of the Pauli-twirled $i$Toffoli, the reference, and the $i$Toffoli gates are obtained as $97.51(2)\%$, $99.24(1)\%$, and $98.26(2)\%$, respectively. The uncertainties represent the fitting errors of the fidelity decay.

\begin{figure*}[h]
    \centering
     \includegraphics[width=1\columnwidth]{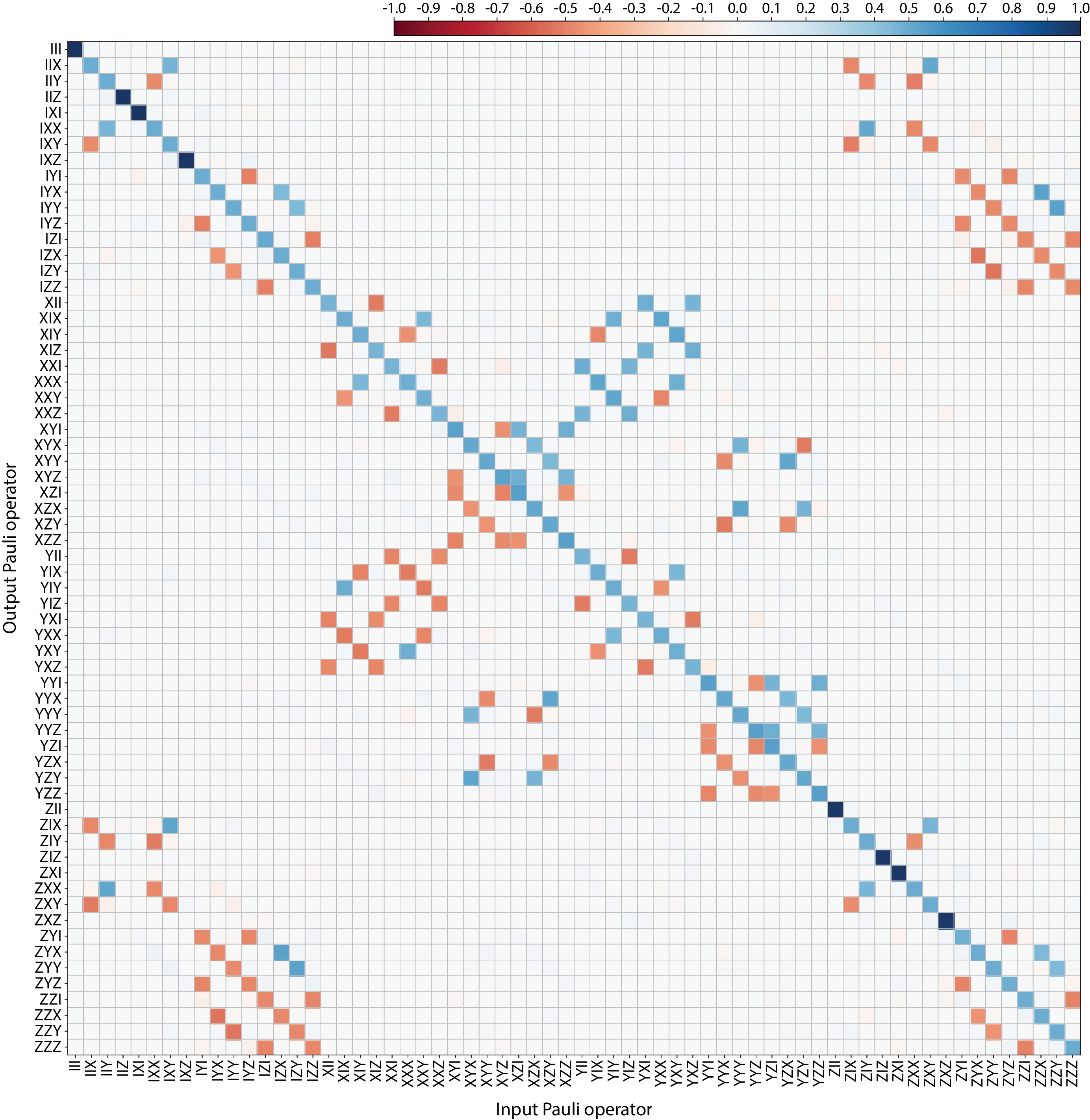}
    \caption{Experimentally reconstructed Pauli transfer matrix of a 353-ns iToffoli gate, $\mathcal{E}_\text{exp}$, via maximum likelihood estimation method. The readout errors are corrected based on the readout fidelities. The process fidelity is estimated to be 97.1(8)\% with respect to the ideal gate.}
    \label{fig:figS2}
\end{figure*}

\begin{figure*}[h]
    \centering
     \includegraphics[width=1\columnwidth]{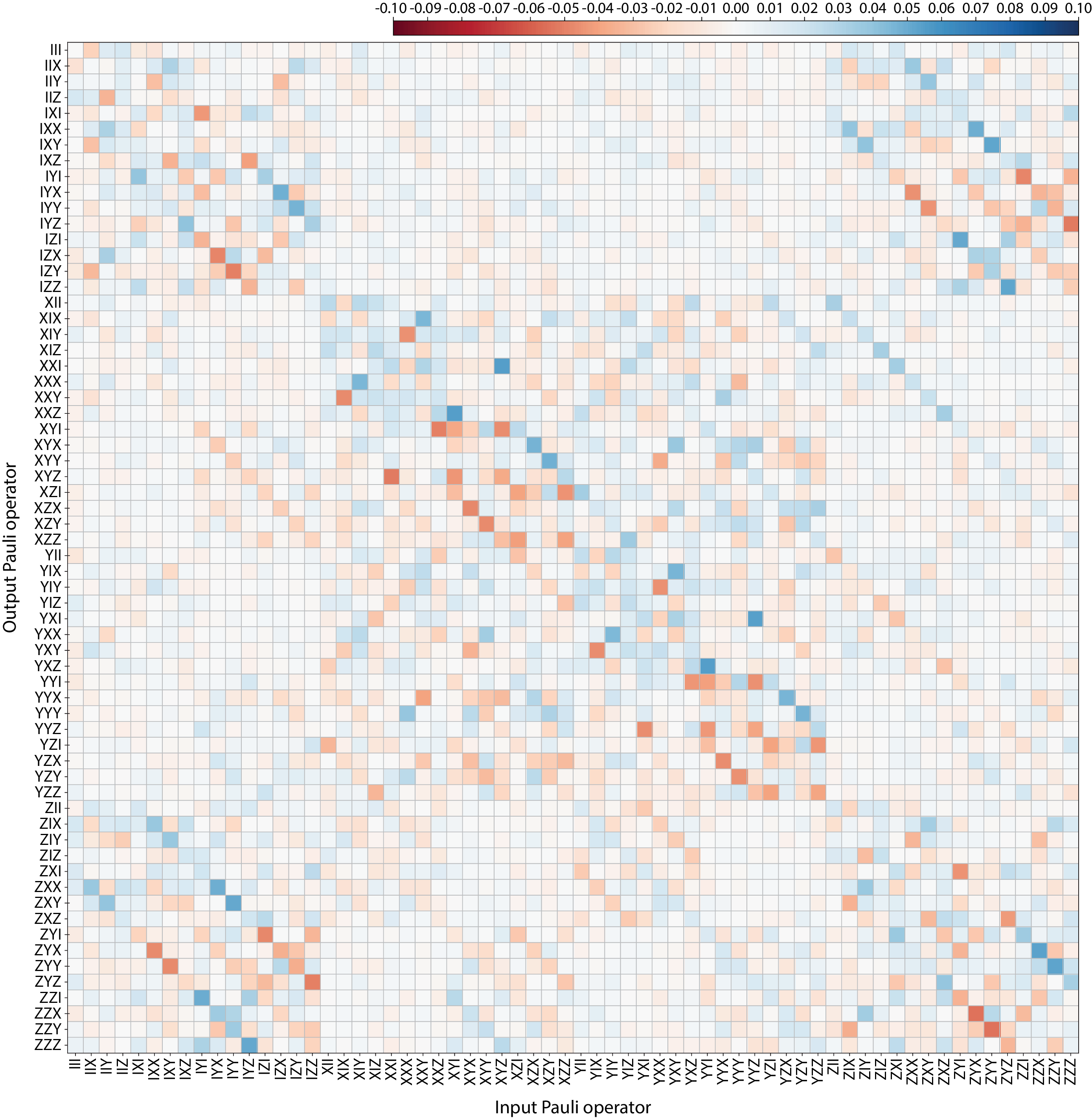}
    \caption{The difference between the ideal $i$Toffoli gate's Pauli transfer matrix (PTM)  and the experimentally reconstructed PTM in Fig.~S\ref{fig:figS2}, $\mathcal{E}_\text{ideal}-\mathcal{E}_\text{exp}$. The largest element difference is 0.055. Note that the color bar is rescaled to a range between -0.1 to 0.1 for clarity.}
    \label{fig:figS3}
\end{figure*}

Second, in order to see the quantum operation of a 353-ns $i$Toffoli gate more explicitly, the Pauli transfer matrix (PTM) is reconstructed. To this end, we prepare 64 three-qubit input states by applying single-qubit operations $\{\text{I}, \text{X}_{\pi/2}, \text{X}_{\pi}, \text{Y}_{\pi/2}\}$ for each qubit. Then, we measure the output states and reconstruct the PTM from
the data via maximum likelihood estimation method~\cite{Chow12}. The readout errors in the data are corrected based on the readout fidelities~\cite{Nachman20}. Figures~S\ref{fig:figS2} and~S\ref{fig:figS3} show the reconstructed PTM, $\mathcal{E}_\text{exp}$, and the difference with the ideal PTM, $\mathcal{E}_\text{ideal}$. The process fidelity is evaluated to be $\mathcal{F}_{\text{PTM}}=\text{Tr}[\mathcal{E}^{T}_{\text{ideal}}\mathcal{E}_\text{exp}]/d^2=97.1(8)\%$ where $d=8$ is the Hilbert space dimension~\cite{Nielsen02}. The uncertainty is obtained by performing 1000 Monte Carlo simulation runs taking into account statistical errors.

\newpage

\newpage

\newpage
\section{Supplementary note 3 - Decoherence under \MakeLowercase{i}Toffoli drive}
As discussed in the main text, decoherence of qubits is one of the main error sources in our $i$Toffoli gate, along with geometric phase errors. When implementing the $i$Toffoli gate, we drive control qubits at off-resonance frequency and the drive shifts the qubit frequencies. The frequency shift could induce faster energy relaxation (lower $T_1$) of control qubits if the qubit frequencies come into resonance with strongly coupled two-level systems~\cite{Carroll21}. On the other hand, a faster dephasing effect (lower $T_2$) under off-resonant drive has also been experimentally observed ~\cite{Brad21}.

\begin{figure*}[h]
    \centering
     \includegraphics[width=0.8\columnwidth]{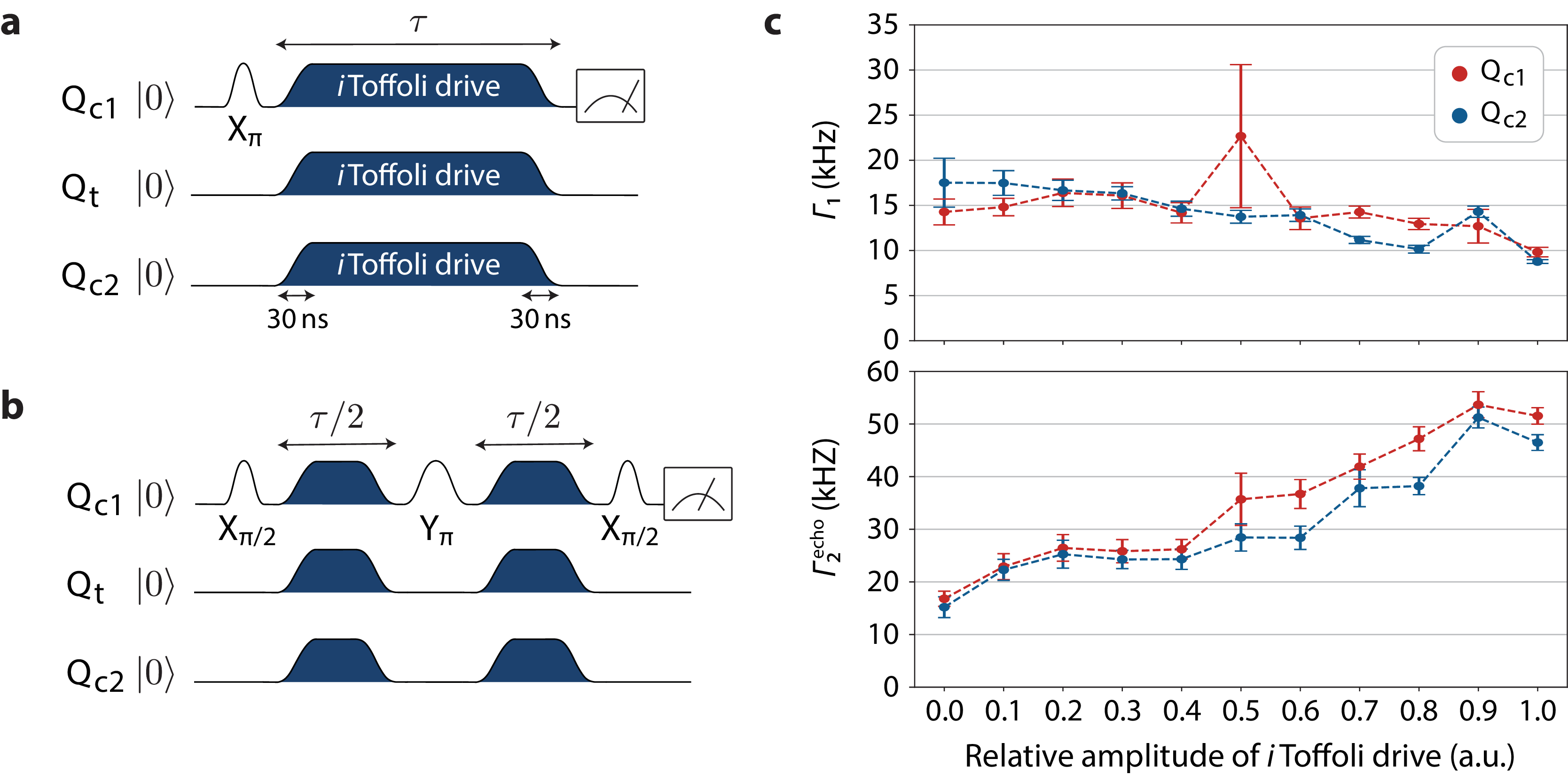}
    \caption{\textbf{Decoherence of control qubits under \textit{i}Toffoli drive.} Pulse sequences used to measure, \textbf{a}, the energy relaxation rate $\varGamma_1=1/T_1$ and, \textbf{b}, the Hahn-echo dephasing rate $\varGamma_2^\text{echo}=1/T_2^\text{echo}$  of the first control qubit Q$_\text{c1}$. The decoherence rates of the second control qubit Q$_\text{c2}$ are separately measured in the same manner. \textbf{c}, The measurement results of $\varGamma_1$ and $\varGamma_2^\text{echo}$ versus relative drive amplitudes. The error bars represent statistical fluctuations.}
    \label{fig:figS5}
\end{figure*}

To investigate these decoherence effects under the $i$Toffoli drive, we measure the energy relaxation rate $\varGamma_1=1/T_1$ and the Hahn-echo dephasing rate $\varGamma_2^\text{echo}=T_2^\text{echo}$ of control qubits as functions of a relative $i$Toffoli drive amplitude. Figures S\ref{fig:figS5}\textbf{a} and \textbf{b} depict the pulse sequences for the measurements, respectively. From the experimental results shown in Fig. S\ref{fig:figS5}\textbf{c}, we observe that $\varGamma_2^\text{echo}$ of Q$_\text{c1}$ and Q$_\text{c2}$ gradually increases as relative drive amplitude increases, whereas $\varGamma_1$ is relatively independent to the drive amplitude except the peaks at relative amplitudes of 0.5 and 0.9. The peaks are suspected to be due to frequency collisions caused by frequency shifts. Note that the $i$Toffoli gates with gate durations of 482, 353, 280, and 243 ns, in Fig. 4 of the main text, are implemented at relative amplitudes of 0.2, 0.3, 0.4, and 0.5, respectively.

